\documentclass[11pt]{iopart}
\usepackage{iopams}  
\usepackage[ampersand]{easylist}
\usepackage{graphicx}
\usepackage{color}

\begin{document}

\title[The Search for Dark Matter]{The Search for Dark Matter}

\author{Laura Baudis}

\address{Physik Institut, University of Zurich, Winterthurerstrasse 190, 8057 Switzerland}
\ead{laura.baudis@uzh.ch}

\begin{abstract}

The dark matter problem is almost a century old. Since the 1930s evidence has been growing that our cosmos is dominated by a new form of non-baryonic matter, that holds galaxies and clusters together and influences cosmic structures up to the largest observed scales. At the microscopic level, we still do not know the composition of this dark, or invisible matter, which does not interact directly with light. The simplest assumption is that it is made of new particles that interact with gravity and at most weakly with known elementary particles.  I will discuss searches for such new particles,  both space- and Earth-bound including those placed in deep underground laboratories. While a dark matter particle hasn't been yet identified, even after decades of concerted efforts, new technological developments and experiments have reached sensitivities where a discovery might be imminent, albeit  certainly not guaranteed. 

\end{abstract}

\section*{Introduction}

What is our cosmos made of? We do not know. For millennia, we have experienced the Universe through light. Light in the visible region, seen by the unaided eye. Starting with Galileo, we used telescopes to observe the night sky, and rather quickly a new cosmology - which placed the Sun and not the Earth at the centre of our Universe  - emerged. In the following centuries, we  built ever more sophisticated telescopes that can map not only the solar system, but our own galaxy and the more distant Universe in many different photon wavelengths, from the radio regime to the very high energy gamma radiation.  We built large-scale detectors that can observe  charged cosmic particles and even neutrinos, up to energies which are ten million times higher than those produced in the most powerful accelerators on Earth. And most recently, a so far quiet window on our Universe was revealed with the measurement of gravitational waves from black hole and neutron star mergers, the latter accompanied by observations in almost all wavelengths of light. These measurements have further confirmed that our standard gravitational theory, based on Einstein's general relativity, is correct, and have shown for the first time that gravitational waves, the ripples in spacetime, propagate with the speed of light.
%(to within a tenth of a trillionth of a percent).

Nevertheless, one of the greatest mysteries is yet to be solved - the identification of the nature of dark matter, the invisible substance that governs the masses of galaxies, clusters and shapes the observed cosmic structures on largest scales. This problem poses an enormous challenge but also a golden opportunity for researchers in a variety of fields, who joined efforts  to solve one of the greatest enigmas of the twenty-first century.

\section*{Evidence for Dark Matter}

About 80 years ago, the Swiss-American astronomer Fritz Zwicky operated a telescope at the Mount Wilson Observatory in California to measure the radial velocities of galaxies (then called {\sl{nebulae}}) in the Coma cluster, a rich cluster of galaxies  about 320 million light years away from Earth \footnote[1]{One light year is the distance travelled by light in one year: $\sim$9.5$\times$10$^{12}$\,km.}. He found surprisingly large velocity dispersions, indicating that the cluster density was much higher than the one derived from luminous matter alone, with a mass-to-light ratio of about 500\footnote[2]{By modern standards, there were several issues with Zwicky's original estimate, including the wrong value of the Hubble constant, which is the current expansion rate of the Universe, crude estimates for the luminosity, the cluster radius and poor statistics; nonetheless, his main result survived, and the current value for the mass-to-light ratio  (the ratio of total mass to luminosity) of galaxy clusters incidentally asymptotes 400.}. He concluded that "should this turn out to be true, the surprising result would follow that dark matter [{\sl dunkle Materie} in German] is present in a much higher density than radiating matter" \cite{Zwicky:1933gu}.  

In the 1970's, Vera Rubin {\sl et al.} and Albert Bosma measured the rotation curves of spiral galaxies with optical images and also found evidence for a {\sl missing mass} component: the rotation curves remained flat, and did not decline in their outer reaches, as predicted by the distribution of luminous matter alone.  By the early 1980s, evidence for this non-luminous matter was firmly established on galactic scales, once rotation curves of galaxies well beyond their optical radii were measured in radio emission, using the 21\,cm line of neutral hydrogen gas \cite{Bosma:1981zz, Rubin:1985ze}. 

In the following decades, the existence of dark matter was inferred using increasingly diverse and precise observations, as well as numerical simulations of galaxy and cosmic structure formation. On galactic and cluster scales, it relies on more accurate measurements of galactic rotation curves,  and on complementary methods to map the total mass of clusters: measurements of orbital velocities  of individual galaxies, mass determinations using the gravitational lensing effect\footnote[3]{Einstein first discussed the gravitational lensing by stars in 1936, while Zwicky proposed gravitational lensing by galaxies in 1937, a subject which once again was ignored until much later, in the 1970s, with the first galactic lens being discovered in 1979.}, and via  the distribution of the hot, X-ray emitting gas, which is a measure of the total gravitational potential.  

The study of the cosmic microwave background (CMB), an almost-uniform background of microwaves reaching us from all directions in the sky, brings one of the strongest indications of dark matter, giving us a precise map of the density of matter in the early Universe \footnote[4]{This radiation was emitted about 380'000 years after the Big Bang, when the Universe was filled with a hot, ionised gas. Once electrons combined with protons to form neutral hydrogen atoms,  photons could start to travel freely through space.}. The observed tiny fluctuations, at the level of one part in 100'000 when the radiation was emitted, grew with time due to gravity, eventually providing the seeds for galaxies and galaxy clusters that we observe today. The shape of the amount of fluctuations, the so-called power spectrum, in the CMB temperature spectrum at different angular scales,  is determined by the oscillations in the hot gas in the early Universe,  while the resonant frequencies of these oscillations and their amplitudes depend on the composition. By analysing the power spectrum cosmologists were able to determine that our Universe is spatially flat, accelerating, and composed of 5\% baryonic matter, 27\% cold dark matter (CDM) and 68\% dark energy ($\Lambda$) \cite{Ade:2015xua}. Baryonic matter is the ordinary matter that forms planets, stars, galaxies, and the diffuse gas in between. The dark energy, a sort of cosmic fluid that pushes structures apart and thus counteracts gravity, is independently observed through its effect on the expansion of our Universe. An accelerated expansion was inferred from comparing the observed redshift and luminosity distances of distant {\sl standard candles}, in the form of type Ia supernovae \cite{Frieman:2008sn}, and so far is consistent with a cosmological constant, meaning that the density of dark energy remains constant in time. Finally, numerical simulations based on this  standard model of cosmology,  the so-called $\Lambda$CDM model, successfully predict the formation of observed large-scale structures, and of galaxies \cite{Frenk:2012ph}. 

In spite of the vast body of evidence, the dark matter is observed merely indirectly: through its gravitational influence on luminous, baryonic matter.  Our understanding  of its amount and distribution on various astronomical scales, including our own galaxy has been growing over the years, but we are yet to answer the most fundamental question: what is the nature of dark matter? We know that it is 
nonbaryonic, because the measured abundance of light elements produced in the primordial nucleosynthesis\footnote[5]{The primordial nucleosynthesis describes the formation of the light elements $^2$H, $^3$He, $^4$He and $^7$Li from about 10 s to 20 min after the Big Bang. The resulting abundances once nucleosynthesis ends, which are compared with measured ones in astronomical objects where little stellar nucleosynthesis occurred, depend on the baryon-to-photon number ratio, and are thus a measure of the amount or baryons, or ordinary matter, in the early Universe.} is much smaller than the total density of matter, a result which was confirmed with higher precision by measurements of the CMB.  In addition, searches for faint astronomical bodies such as stellar remnants, low mass stars, planets and black holes via gravitational microlensing have shown that such Massive Compact Halo Objects (MACHOs) can not substantially contribute to the  Milky Way's dark halo \cite{Moniez:2010zt,Strigari:2013iaa}.

The flat galactic rotation curves and the systematics of galaxy properties can  be exceedingly well explained by modifying the laws of gravity at large scales, e.g., in the framework of modified newtonian dynamics, MOND, as proposed by Moti Milgrom in 1983, or more recent, relativistic versions thereof. These radical alternative theories to dark matter have failed to explain clusters and the observed pattern in the CMB. In addition, they predict that photons and gravitational waves should travel on different space-time paths (geodesics), implying a delay of about 800 days between the  arrival time of gravitational and electromagnetic waves over a distance of 130 million light years. This prediction is in stark contrast  to the observed delay of 1.7\,s in the recent detection of gravitational waves and gamma-ray signals from the merging neutron star event called { GW170817}. 

\section*{Dark Matter Candidates}

Perhaps the simplest assumption is that, at the microscopic level, dark matter is made of a new kind of elementary particle, yet to be discovered.  Instantly, additional questions arise: what are the properties of this particle, namely its mass,  its spin and its interaction probability (or cross section) with normal matter? Is it  just one particle species, or many? After all, the visible, normal matter is composed of different kinds of elementary particles, do we expect the {\it dark sector} to be different? Are these particles stable, or very long-lived? While we have no answers to these questions, we know that such particles would not interact via the electromagnetic force (as far as we can tell with current telescopes and experiments) and would be much longer lived than the age of our Universe, which is 13.8 billion years. Possible masses span an enormous range, from about 10$^{-22}$\,eV  to 10$^{15}$\,GeV, or 46 orders of magnitude\footnote[6]{The electron-volt (eV) is the characteristic energy scale of atomic physics processes, for example the ionisation energy of the hydrogen atom is 13.6\,eV. The kilo electron-volt (keV) is the energy scale of X-rays, while mega electron-volt (MeV) is typical for nuclear physics processes. The giga electron-volt (GeV) scale is characteristic for the rest mass energy of a proton, the hydrogen atom nucleus, while at the LHC, the proton-proton collider near Geneva, protons are accelerated up to kinetic energies of 7000\,GeV.}.
%the highest observed energies on Earth, from cosmic rays, are roughly  $10^5$\,GeV.}.
In terms of the interaction strength with known, standard model particles, the allowed range is even larger, extending roughly 60 orders of magnitude. Clearly no single experimental technique can cover such a vast parameter space. In this space, two islands clearly stand out: so-called weakly interacting massive particles (WIMPs) and axions, for these particles were not specifically invented to solve the dark matter problem.
If WIMPs exist, and were in thermodynamic equilibrium with the rest of the particles in the hot plasma of the early Universe, their abundance today is close to the measured abundance of dark matter from astrophysical and cosmological observations ($\Omega_{DM} \approx 0.27$) if their velocity-averaged annihilations cross section is around the electroweak scale, namely about $3 \times 10^{-26} {\rm cm}^3 {\rm s}^{-1}$\footnote[7]{The abundance of dark matter is given in terms of the critical density $\rho_c = 3 H_0^2 M_{Pl}^2/8\pi = 1.88\times10^{-29}$\,g\,cm$^{-3}$ with the Hubble constant $H_0 \simeq 70\,$km/(s Mpc) and the Planck mass $M_{Pl}$=$10^{19}$\,GeV, namely $\Omega_{DM} = \rho_{DM}/\rho_c$. The critical density $\rho_c$ corresponds to about 6 hydrogen atoms per cubic meter of space. }. This value serves as a benchmark for some of the dark matter detection methods, so-called {\sl indirect searches}, that I will discuss later on.

None of the particles in standard model  meets the requirements to be a dark matter candidate. In the 1980s neutrinos were considered as a serious possibility, for they do not interact with light and we know that these elusive particles exist.  However, we know from direct experiments and cosmological observations that neutrinos are very light, lighter than 2\,eV, and their velocity would be large enough to affect structure formation. Neutrinos are an example for the so-called {\sl hot dark matter}, a scenario in which superclusters  would form first, and later fragment to galaxies, in contrast to observations \cite{Frenk:2012ph}.  On the other hand, WIMPs are candidates for {\sl cold dark matter}:  they would have been slow-moving by the time they decoupled from the hot plasma in the early Universe, allowing smaller structures such as galaxies to form first, and later merge to clusters, and superclusters, in agreement with observations and numerical simulations of structure growth, which well-reproduce the large-scale structures observed today. WIMPs are predicted by various theories that address theoretical problems within the standard model  itself, examples being the lightest supersymmetric particle in supersymmetric extensions\footnote[7]{Supersymmetry is a spacetime symmetry, that proposes a relationship between particles with half-integer spin, or fermions, and particles with integer-spin, or bosons. Examples for fermions are electrons and protons, examples for bosons are photons and gluons.},  or the lightest excitations (so-called Kaluza-Klein particles) in theories with extra dimensions  of space.

The second island candidates in our vast dark matter candidate space is made of axions, or more generally, axion-like particles.  Axions were first introduced as a solution to the strong-CP problem\footnote[8]{C stands for charge conjugation, and P for parity transformation; the CP operation is not conserved in the weak interaction.} in quantum chromodynamics (QCD), the theory of strong interaction which binds neutrons and protons inside atomic nuclei.  While QCD is a remarkably successful theory, it also predicts a dipole moment for the neutron, which is a measure for its distribution of positive and negative charge, that is about 10 orders of magnitude larger than existing experimental limits, $d_n< 2.9 \times 10^{-26}$\,e\,cm. To solve this problem, Peccei and Quinn introduced a new global symmetry of the theory which is spontaneously broken below an energy scale called $f_a$ \cite{Peccei:1977hh}. Weinberg and Wilczek then realised that there must be a new particle (a so-called Nambu-Goldstone boson), which they called the axion, associated with the global symmetry breaking. The axion mass turns out to be inversely proportional to the symmetry breaking scale, $f_a$, which is arbitrary, for all values solve the strong CP problem. For instance, if   $f_a$ is between 100\,GeV and 10$^{19}$\,GeV, the axion mass is between 1\,MeV and 10$^{-12}$\,eV, which spans a rather large range.  Initially, Peccei and Quinn proposed  a symmetry breaking scale around the electroweak scale ($v_{weak}$ = 247\,GeV), leading to an axion mass  around 200\,keV. This axion was soon excluded by direct experimental searches and astrophysical arguments, that considered the effects of axion emission on the evolution of red-giant stars. So-called invisible axion models, with  $f_a \gg v_{weak}$ are still viable \cite{Preskill:1982cy}, as they evade all the existing experimental bounds.  Although they are very light, axions would also constitute {\it cold dark matter}, because they were produced non-thermally in the early Universe. Their masses are restricted to a narrower range, from 1\,$\mu$eV to 3\,meV by laboratory searches and from various astrophysical observations \cite{Agashe:2014kda}.  In more general models, where axion-like particles are predicted to arise in low-energy effective field theories emerging from string theory,  the two parameters, namely the particle mass and its decay constant are independent of one another, opening up an even larger parameter space for searches.

\section*{Detection of Weakly Interacting, Massive Particles}

The detection of WIMPs has spawned many imaginative solutions, but these can be roughly classified as searches for galactic dark matter particles by  so-called direct and indirect methods, and production at high-energy colliders. Since the masses of {\small WIMPs} are predicted to be in a range from a few\,GeV, hence a few times the mass of a hydrogen nucleus, to about 10$^5$\,GeV, experiments must cover at least five orders of magnitude \cite{Baudis:2015mpa}.  Direct detection experiments search for a tiny scattering signal between a WIMP and an atomic nucleus in experiments with an ultra-low background noise from natural radioactivity, operated deep underground. Indirect detection experiments are poised to observe familiar particles, such as neutrinos, gamma rays, antiprotons and positrons above the astrophysical background, born when dark matter particles in our galaxy meet and annihilate. Finally, experiments at colliders such as the LHC look for the emergence of new, dark matter particles in high energy collisions, where the kinetic energy of the colliding particles is transformed into the mass of the new, heavy particle. Since this particle will escape any detector unnoticed, the tell-tale signature is missing energy, together with at least one trace produced by familiar particles such as quarks, gluons, photons or Z-bosons, that is required for recording the event. After intense efforts by very large collaborations, there is no evidence for dark matter or any other type of new particle, not part of the standard model, from the LHC. Nonetheless, more high-quality data has been, and will be collected, and the search is ongoing.

The idea that WIMPs can be detected by their elastic collisions with atomic nuclei in Earth-bound detectors goes back to Goodman and Witten \cite{Goodman:1984dc}, following an earlier suggestion of Drukier and Stodolsky in 1983  to detect solar and reactor neutrinos by exploiting their elastic neutral-current scattering of nuclei.   Interestingly, such low-energy, neutrino-induced scatters were for the first time detected in 2017 with a 14\,kg CsI scintillating crystal operated near a strong neutrino source at Oak Ridge National Laboratory in USA, by the Coherent experiment. The initial study was soon extended by Drukier, Freese and Spergel \cite{Drukier:1986tm} to include various cold dark matter candidates and to show that the Earth's motion around the Sun, together with the Sun's motion around the galactic centre, induce an annual modulation in the expected signal rate. These and subsequent studies  laid the foundations for an assortment of experimental approaches that by now  have sufficiently matured such that one, or even multiple discoveries might be around the corner.

The first detectors that looked for WIMP dark matter were high-purity Ge crystals enriched in $^{76}$Ge, designed to search for a very rare nuclear decay process, called neutrinoless double beta decay. This decay can only happen if neutrinos are their own antiparticle, a question which is still open today.  These early experiments, one of which was the Heidelberg-Moscow experiment that I was part of,  were  able to exclude a heavy, right-handed Dirac neutrino as a dark matter candidate, as well as the cosmion, a particle with a mass of a few GeV and a cross section of 10$^{-36}$\,cm$^2$ that had been postulated to solve both, the dark matter and the solar neutrino problem. The technological progress during the following two decades was tremendous, but today's direct detection experiments still face two major challenges: the very small energies released in a collision with an atomic nucleus in a detector and ultra-low scattering rates.  While each challenge has been solved on its own (small energies are detected with bolometers, or phonon-based detectors, while ultra-low backgrounds have been achieved in large detectors looking for solar and atmospheric neutrinos), the main difficulty has been to combine both features, namely to build large detectors, with ultra-low backgrounds and energy thresholds of a few keV.

The small required energies, well below 100\,keV come from the fact that WIMPs move slowly relative to the detector, with a mean velocity of about 220\,km/s. This velocity comes from our movement through the halo of dark matter particles. For light particles with masses of 1-10\,GeV, the recoil energy is even lower, at most a few keV. To design and eventually build such an experiment, we must estimate the expected scattering rate: it depends not only on the particle mass, but also on the density of dark matter particles in the vicinity of the Sun, which is measured to be  around 7$\times$10$^{-25}$\,g/cm$^3$ (or 0.4\,GeV/cm$^3$ in units familiar to particle physicists), \cite{Read:2014qva} on their mean velocity and velocity distribution, on the mass and number of  target nuclei in the detector and on the interaction strength (or cross section).   

For a particle mass of 100 times the mass of the proton, this implies a WIMP flux onto the Earth of about 100'000\,cm$^{-2}$s$^{-1}$ \cite{Baudis:2012ig}. This flux allows us to in principe detect WIMPs as they pass through and scatter elastically off atomic nuclei in an Earth-bound detector.   Because of the unknown scattering cross section, the expected event rates  range from about 10 to much less than 1 event per tonne of detector material and year.   This is an extraordinarily  small rate, implying that we would need to measure for many years even with a ton-scale detector, to accumulate a significant number of signal events.
To maximise the probability of a detection, experiments must thus feature a very low energy threshold, an ultra-low background noise and a target mass which is a large as possible. 
A broad range of techniques is used to search for the tiny  energy of the scattered  atomic nucleus, which in a detector is converted into a measurable signal, usually ionisation, scintillation light or heat. Among those with highest sensitivities are bolometric crystals operated at sub-Kelvin temperatures, detectors using liquified  xenon and argon, and superheated liquid detectors  \cite{Baudis:2012ig}. In addition, several detectors that are capable of measuring the direction of the recoiling nucleus are in R\&D stage. 

Starting from the first experiments dedicated to dark matter searches, the main background sources were the radioactivity of detector construction materials (mainly $^{238}$U, $^{232}$Th and $^{40}$K) and the dark matter target itself,  and cosmic rays. To shield from the latter, all experiments are operated deep underground, using the Earth as a filter: more than a dozen of laboratories are located on our planet, some in former or even operating mines, others especially built off road or railway tunnels through mountains. The experiments I work on are at  the Gran Sasso Underground Laboratory (LNGS) in Italy, in a tunnel  shielded by about 1400\,m of rock of the Corno Grande. The presently deepest underground laboratory, JinPing, is located in China, in a service tunnel parallel to the water ducts of a hydroelectric power station. It is under a mountain overburden of 2400\,m, reducing the cosmic muon flux to only 2 muons per square meter and 10 days, while the flux at LNGS is 1 muon per square meter and hour.

As detectors finally reach the tonne and multi-tonne scale, the radioactivity from detector materials becomes less relevant (assuming that the dark matter target material itself can be purified to a very high degree), and the main and possibly irreducible backgrounds will come from neutrino interactions. This seemed rather improbable twenty years ago, when I started to work in this field.  Low-energy neutrinos present in large fluxes  from the pp and $^{7}$Be reactions in the Sun  give rise to neutrino-electron scatters, while the ultimate background might come from neutrino-induced nuclear recoils from coherent neutrino-nucleus scatters. The scattering of $^{8}$B solar neutrinos can mimic WIMPs with masses around 5-6\,GeV, while neutrinos from the atmosphere and diffuse supernova neutrinos can mimic a WIMP-signal for masses  above 10\,GeV/$c^2$. However, the sensitivity of experiments must be increased by more than a factor of 100 until such neutrinos will become visible. They may even be promoted from background to an actual signal, for there are relevant neutrino physics questions that can be studied, e.g. the measurement of the pp-neutrino flux with 1\% or better precision, and observation of non-standard coherent neutrino-nucleus scatters.the 

In spite of over two decades of searches, there is no convincing evidence for WIMP dark matter from any direct detection experiment. The mass region below 6\,GeV is studied with detectors with very low energy thresholds and/or lighter target nuclei,  such as those detecting phonons, and light or charge (CRESST, EDELWEISS, SuperCDMS) or based on CCD technology to record the particle tracks (DAMIC). The higher mass region above 6\,GeV is probed by experiments using liquefied argon and xenon, with very low background rates and much higher target masses. In particular detectors using liquid xenon have successfully built and operated  large time projection chambers (LUX/LZ, XENON1T/nT and PandaX) with very good 3 dimensional position resolution, and these provide the tightest constraints on the interactions strength of WIMPs with nuclei.  

The XENON1T detector at the Gran Sasso laboratory operates a total of 3.2\,t of liquid xenon, 2\,t of which are in the time projection chamber, constantly observed by 248 photomultipliers with a high quantum efficiency for the xenon scintillation light in the vacuum ultraviolet region (178\,nm). A uniform electric drift field allows for the drift and extraction of electrons into a vapour phase above the liquid, where these are amplified and produce a secondary, delayed scintillation signal. XENON1T has been taking data for a long science run since February 2017. While the data is constantly being analysed,  the region where dark matter signals are expected is masked, or blinded. This virtual signal box will be opened in early 2018, when the analysis will be ready. The next-generation DARWIN detector, to operate 50 tonnes of liquid xenon, is designed to probe the entire parameter space for WIMPs, until the neutrino-induced background will eventually start to become visible.

Dark matter particles might also be detected by observing the radiation produced when they annihilate, and this is called the indirect method. 
Although after the decoupling from the rest of the particles in the early Universe the dark matter pair annihilation becomes largely suppressed, the self-annihilation process that was relevant then could still be significant today. Hence dark matter particles would annihilate into standard model particles,  which then decay and give rise to familiar products that could potentially be observed above the standard, astrophysical background.  Examples are energetic cosmic rays, such as neutrinos, gamma rays, antiprotons or positrons.

Because the flux of annihilation products is proportional to the number density of dark matter particles squared,  those regions in our galaxy with largest densities are the most interesting to observe. 
We would thus expect high energy neutrinos from the Sun's or Earth's cores, gamma-rays from the Galactic Centre or the region around it, from sub-halos of the Milky Way or even from satellites such as the dwarf spheroidal galaxies. Another signature would be positrons and antiprotons from the galactic halo. 

The predictions of actual expected fluxes are far from trivial: these obviously depend on the particle mass, annihilation strength, and dark matter density profile. They also depend on the presence of dark matter sub-structure
and the galactic cosmic ray diffusion model, the latter being particularly relevant for the propagation of charged particles such as positrons and antiprotons. One of the main challenges is to model correctly  the expected backgrounds from standard astrophysical processes, as these could mimic  a potential dark matter signal.

Several astrophysical observations had been interpreted as signatures for dark matter annihilation in our galaxy during the last ten years. Two of the most prominent were an unexpected rise in the fraction of positrons to electrons measured first by the PAMELA satellite and more recently by the AMS-02 experiment  on the international space station, and an excess of 1-3\,GeV gamma rays from the region surrounding the Galactic Centre  observed in the data of the Fermi satellite.  The observed rising positron fraction with positron energy implies a new source of positrons: these could be ejected by pulsars,  which are rotating, highly magnetised neutron stars,
or be due to dark matter particle annihilation. The latter possibility now seems unlikely, and is also in strong tension with measurements of  the cosmic microwave background radiation by the Planck satellite. WIMP annihilation at early times would give rise to a sufficient number of energetic particles to impact the observed anisotropies. 
 
 The second subject which underwent intense debates was the observed gamma-ray excess emission in the inner galaxy, up to galactic latitudes of about 10 degrees, the so-called Galactic Centre excess. It shows a very uniform spectrum and an approximately spherically symmetric distribution.  This signal morphology is just what would be expected from dark matter annihilation.   If the excess over the astrophysical background  is interpreted as due to dark matter annihilation, the mass of the particle would be in the $10 - 50$\,GeV range, and the annihilation strength  in the correct range expected for WIMP dark matter.  There is quite some support however, that the signal is due to discrete sources, and not from  a smooth, dark matter distribution - where the sources could be a new  population of pulsars. In addition, the Fermi collaboration does not observe a gamma-ray excess in Milky Way satellites (so-called dwarf spheroidal galaxies),  which are some of the most dark matter dominated galaxies known.

Even under the daring assumption that these anomalies are due to dark matter,  different particles with masses masses in the range $10-50$\,GeV, or of several TeV  would be required to explain the data. There is some hope that new data, not only from the operation of Fermi and AMS-02, but also from the current generation of atmospheric Cherenkov detectors 
and the future Cherenkov Telescope Array, as well as from existing and future neutrino experiments that look for neutrinos from dark matter annihilation in the Sun, will shed some light into the problem.

\section*{Axion Detection}

Searches for axions take advantage of the predicted minuscule coupling of this particle to  photons, electrons, as well as to protons or neutrons. For standard axions, motivated by the strong CP problem in QCD, the coupling strengths to standard model particles are inversely proportional to the so-called called the axion decay constant $f_a$, or equivalently to the axion mass $m_a$, which is the only free parameter of the model.  The mass of the axion is related to the mass of the pi meson (or pion), $m_{\pi}$ = 135\,MeV, a particle made from a quark-antiquark pair  involving the lightest quarks up and down, which decays into two photons, $m_a f_a \approx m_{\pi} f_{\pi}$, with the measured pion decay constant $f_{\pi} \approx$ 92\,MeV. Thus, the smaller the axion mass, the weaker its couplings to ordinary particles.  

The larger part of axion searches exploits the predicted coupling to two photons, which implies that the axion could convert into a photon in an external electric or magnetic field.  In light-shining-through-wall experiments laser photons that propagate through a transverse magnetic field could convert into axions, which are to pass an optical barrier, an opaque "wall" and then reconvert into photons in a second magnetic field. The highest sensitivity is currently reached by the ALPS-II experiment at DESY in Hamburg. Axion helioscopes, such as the CAST experiment at CERN and the proposed IAXO, look for axions that might be produced in the hot solar interior, with an average energy of 4.2\,keV, the temperature at the core of the Sun. These could convert into X-rays in a magnetic field within a dipole magnet (CAST uses an old, refurbished magnet from the LHC) oriented towards and tracking the Sun, and detected with X-ray detectors placed at both ends of the magnet. 

Axion haloscopes, the prime example being the ADMX experiment at the University of Washington,  were built to search for the cold dark matter axion.  Assuming a local dark matter density of 0.4\,GeV\,cm$^{-3}$, an axion mass of 0.5\,$\mu$eV and particle velocities of one thousandth of the velocity of light, the axion flux on Earth is roughly 10$^{21}$ cm$^{-2}$s$^{-1}$. 
While the lifetime of the axion is extremely long,  the axion can convert to photons in a static, external magnetic field.  In the scheme proposed by Pierre Sikivie in 1983 \cite{Sikivie:1983ip}, the axion-photon conversion is enhanced in a microwave cavity permeated by a strong magnetic field when the resonant frequency of the cavity equals the axion rest mass $m_a$. The resonant cavity is tuneable, hence a range of axion masses can be scanned by changing the frequency, where the axion signal is detected by observing the proper modes at a given frequency. The new ADMX experiment started to take data in 2016 and it will test a sizeable fraction of the predicted parameter space for the QCD axion as a dark matter candidate.  In recent years, many new ideas were put forward to detect axion-like-particles, for which  $m_a$ and $f_a$ are not necessarily related, over a large range of masses and coupling strengths. No signs of axions were observed by any of the discussed searches so far, but the quest is ongoing.

\section*{Outlook}

Almost a century after the first discovery of dark (or invisible) matter, its composition at the fundamental level remains an enigma. We live in a vast sea of dark matter, but decades of intensive research have mostly established what it is not. Understanding its nature thus represents one of the formidable challenges of contemporary physics. An enormous effort by theorists, experimentalists and observers alike is being  put forth to better understand and constrain the problem, and there is hope that a discovery might be forthcoming in the not-so-distant future. The continued search for the nature of dark matter presents not only a challenge, but also a golden opportunity: understanding the problem might herald one of the great advances of twenty-first century physics, and  reveal new symmetries, new types of fundamental particles and new forces, while at the same time opening up a new field: dark matter astronomy.

%\newpage

%\section{About the author}
%Laura Baudis' research interests lie at the intersection of particle physics, astrophysics and cosmology.  She builds detectors to discover and hopefully study dark matter particles and to reveal fundamental properties of neutrinos. Baudis received her Ph.D. from the University of Heidelberg in 1999 and went on to become a postdoctoral fellow at Stanford University. At Stanford, she worked on the Cryogenic Dark Matter Search experiment. She moved to the University of Florida, Gainesville, in 2004, as an assistant professor, where she started to work on detectors using liquefied xenon. In 2006 she was awarded the Lichtenberg Professorship for Astroparticle Physics at the RWTH Aachen University, and  in 2007 she joined the Physics Department of the University of Zurich, where she is full professor of experimental physics. Her main research projects are XENON and GERDA, both located at the Gran Sasso Underground Laboratory in Italy. She is a Fellow of the American Physical Society (APS), a member of the CERN Science Policy Committee and an Editor in Chief of the European Physical Journal C. In 2017, she was awarded an ERC advanced grant for her project Xenoscope, which conducts R\&D for the construction of a 50 tonnes liquid xenon time projection chamber for the DARWIN observatory.

\section{References}

\bibliographystyle{iopart-num}
\bibliography{darkmatter_2017}

\end{document}